\documentclass[preprint,superscriptaddress,nofootinbib]{revtex4-1}
\usepackage{amsmath}
\usepackage{amssymb}
\usepackage{graphicx}
\usepackage{mathrsfs}
\usepackage{cancel}
\usepackage{soul}
\usepackage{color}
\usepackage{multirow}
\usepackage{bm}

\newcommand{\beq}{\begin{equation}}
\newcommand{\eeq}{\end{equation}}
\newcommand{\bea}{\begin{eqnarray}}
\newcommand{\eea}{\end{eqnarray}}

\allowdisplaybreaks[1]

\newcommand{\ctilde}{\tilde{c}}
\newcommand{\Arg}{\mathrm{Arg}}

\def\spa#1.#2{\left\langle#1\,#2\right\rangle}
\def\spb#1.#2{\left[#1\,#2\right]}

\begin{document}

\title{Probe CP violation in $H\to \gamma Z$ through forward-backward asymmetry}

\author{Xuan Chen}
\email{xuan.chen@pku.edu.cn}
\affiliation{Institute of Theoretical Physics $\&$ State Key Laboratory of
Nuclear Physics and Technology, Peking University, Beijing 100871,
China}
\affiliation{Center for High Energy Physics, Peking University, Beijing 100871, China}

\author{Gang Li}
\email{gangli@pku.edu.cn}
\affiliation{Institute of Theoretical Physics $\&$ State Key Laboratory of
Nuclear Physics and Technology, Peking University, Beijing 100871,
China}

\author{Xia Wan}
\email{wanxia@snnu.edu.cn}
\affiliation{School of Physics $\&$ Information Technology, Shaanxi Normal University, Xi'an 710119, China}

\date{\today}

\begin{abstract}

We suggest that the forward-backward asymmetry $(A_{FB})$ of the charged leptons in $gg\to H\to\gamma Z\to\gamma \ell^-\ell^+$ process could be used to probe the CP violating $H\gamma Z$ coupling when the interference from $gg\to\gamma Z\to\gamma \ell^-\ell^+$ process is included. With CP violation in $H\gamma Z$ coupling, the interference effect leads to a non-vanishing $A_{FB}$, which is also sensitive to the strong phase differences. The resonant and non-resonant strong phases together make $A_{FB}(\hat{s})$ change sign around Higgs mass $M_H$.
For phenomenology study, we suggest the integral over one-side mass region below $M_H$ to magnify the $A_{FB}$ strength. 

\end{abstract}

\maketitle

\section{Introduction\label{Introduction}}

To explain the observed matter-antimatter asymmetry in the universe,
some CP-violation sources beyond Standard Model (SM) are needed~\cite{Gavela:1993ts,Sakharov:1967dj}.
The Higgs boson discovered five years ago with mass around 125~GeV may provide clues
to study the source of CP violation. Even though the constraint of CP violation from electric dipole moment (EDM) is stringent, it could be evaded in new physics models~\cite{Dekens:2013zca,Inoue:2014nva}. The CP properties of the Higgs boson is studied through $H\to ZZ \to 4l$ decay channel~\cite{Chen:2014gka} where the momenta of four final state leptons could be used to directly construct a CP-odd product.
 The current measurement of $H\to ZZ \to 4l$~\cite{Khachatryan:2014kca} shows the
CP odd/even mixture could be allowed around $\sim$ $40\%$. By contrast, the $H\to \gamma Z $ or $H\to\gamma\gamma$ processes are less considered when probing CP violation since these processes have only three or two final state momenta. 
However, after considering interference effects between Higgs resonance and Standard Model background, several CP-violation observables
could be constructed. Some studies discussed the CP-violation observables in the $H \to \gamma Z  \to \gamma \ell^- \ell^+ $ process:
 the forward-backward asymmetry ($A_{FB}$) of the leptons in $Z$ boson rest frame \cite{Chen:2014ona, Korchin:2014kha}, and the
 angle $\phi$ between the $Z$ production and decay planes
\cite{Farina:2015dua}. We continue the study of interference effects with new CP-violation observables and discuss the phenomenological impact at current and future hadron colliders. 

 The $A_{FB}$ observable reveals the asymmetry of producing CP conjugate final states
 $F$ and $\bar{F}$. If the full amplitude is the sum of two interfering amplitudes, $M=|c_1|e^{i(\psi_1+\xi_1)}+|c_2|e^{i(\psi_2+\xi_2)}$, where
 $\psi_1$, $\psi_2$ are strong phases and $\xi_1$, $\xi_2$ are weak phases,
 the asymmetry depends on the differences of both weak and strong phases:
 \beq
 A=\frac{\sigma(F)-\sigma(\bar{F})}{\sigma(F)+\sigma(\bar{F})}
 \propto|c_1||c_2|\sin(\psi_1-\psi_2)\sin(\xi_1-\xi_2)~.
\label{eqn:A}
 \eeq
 The CP violation could be probed only when both phase differences exist.

 At Large Hadron Collider (LHC), the Higgs boson is mainly produced by gluon fusion through a fermion loop. For $gg \to H \to \gamma Z \to \gamma \ell^- \ell^+ $ process,
  $gg \to \gamma Z  \to \gamma \ell^- \ell^+ $ is an irreducible background process that could have interference effect.
  Ref.~\cite{Farina:2015dua} studied such effect and found that the $\phi$ angle
 between $Z$ production and decay planes could be shifted by a weak phase from CP-violating $H\gamma Z$ coupling, and thus
 is an CP-violation observable.
 Ref.~\cite{Chen:2014ona} studied the $A_{FB}$ through the interference between $H \to \gamma Z \to \gamma \ell^- \ell^+$ and $ H \to \gamma \gamma^{\ast} \to \gamma \ell^- \ell^+$ processes, and estimated that the integrated $A_{FB}$ value is proportional to $\frac{\Gamma_Z}{M_Z}$. However, there is ambiguity about whether the CP-violation is from $H\to\gamma Z$ or $H\to\gamma \gamma$ vertices. If the couplings of both vertices have similar CP violation sources, thus have approximate weak phases, the $A_{FB}$ value would be cancelled severely and become nearly zero.
  Ref.~\cite{Korchin:2014kha} studied the interferences
 not only between $Z/\gamma$ propagators but also from $H\to \gamma \ell^- \ell^+$ at tree level. It showed the $A_{FB}$ distributions that are dependent on
 CP violation parameters in Yukawa couplings. In Ref.~\cite{Li:2015kxc}, the authors studied the CP violation in $Ht\bar{t}$ coupling through $e^+e^-\to H\gamma$ process, which is similar to the inverted process of our current work. However, the definition of $A_{FB}$ in Ref.~\cite{Li:2015kxc} is different from our current work due to different kinematics.

In this article, we revisit the $A_{FB}$ of the charged lepton through interference effect between $gg\to H\to\gamma Z\to\gamma \ell^-\ell^+$ and $gg\to\gamma Z\to\gamma \ell^-\ell^+$ processes with a CP violating $H\gamma Z$ coupling.
 In the first part, we introduce a general model with CP-violation phase factor and the helicity amplitudes involved for both signal and background processes. We also discuss the parity relations of those amplitudes. In the second part, a special frame with kinematic angles is introduced. We make a two-part factorization in such frame for the differential cross section and
 scrutinize the $A_{FB}$ sources. In the third part, we set up numerical simulations using modified \texttt{MCFM} to estimate the $A_{FB}$ values under different mass integral regions. In the last part, we summarize the results and discuss possible future work.

\section{Helicity Amplitudes}
\subsection{Effective operator and CP violation phase $\xi$}
By considering the gluon fusion to Higgs boson, which is the dominant Higgs production channel at hadron collider and the Higgs decay to a photon plus a Z boson, we use the following dimension-5 effective operators to describe the $gg \to H \to\gamma Z$ process,
\begin{equation}
	\label{eqn:HEF}
	\mathcal{L}_{\rm h} = \frac{c}{v}~h\,F_{\mu\nu} Z^{\mu\nu} + \frac{\ctilde}{2v}~h\,F_{\mu\nu}\tilde{Z}^{\mu\nu} + \frac{c_g}{v}~h\,G^a_{\mu\nu}G^{a\mu\nu}~,
\end{equation}
where $F$, $G^a$ denote the $\gamma$ and gluon field strengths, $a = 1,...,8$ are $SU(3)_c$ adjoint representation indices for the gluons, $v = 246$~GeV is the electroweak vacuum expectation value, the dual field strength is defined as $\tilde{X}^{\mu\nu}=\epsilon^{\mu\nu\sigma\rho}X_{\sigma\rho}$, $c$, $\ctilde$ and $c_g$ are complex numbers.

Compare to Standard Model, we add a CP-odd term to study the potential CP-violation effects from $H\gamma Z$ coupling,
which may arise from CP violations in $Hf\bar{f}$ Yukawa coupling, $HVV$ coupling or other new physics. The Higgs boson couples to gluon via effective vertex where the top and bottom quarks are considered to be massive. The masses of the four light quarks are set to zero during our calculation.
The source to bring CP violation in $H\gamma Z$ coupling may also cause CP violation in $Hgg$ coupling. However, it is beyond the scope of the current study.

The $c$ and $\ctilde$ in Eq.~\eqref{eqn:HEF} are complex numbers and have different phases. For the simplicity of current analysis, we make an assumption that their phases are same or have a difference of $\pi$. That is,
\beq
\Arg(c)=\Arg(\ctilde)~\text{or}~\Arg(c)=\Arg(-\ctilde).
\eeq
It is convenient to define
\beq
\xi=tan^{-1}(\ctilde/c),
\label{eqn:xi}
\eeq
which is a CP violation phase (also called weak phase) in helicity amplitudes, in contrast the phase from the complex number $c$ is a strong phase. More details about $\xi$ would be revealed when we discuss parity relation and CP transformation.  
From the definition, $\xi\in (-\pi/2,\pi/2]$. 
When $\xi=0$, it is the SM case; when $\xi\ne 0$, there must exist CP violation and new physics. 
It is worthy to point out that 
 even though $\xi= \frac{\pi}{2}$ or $\xi= \frac{3\pi}{2}$ corresponds to pure CP-odd coupling, it introduces 
CP violation because Standard Model $ggH$ coupling is CP-even. 
As $\xi$ is the only weak phase in our analysis, the CP observable is expected to be proportional to $\sin\xi$, which will be verified later in our analytical calculation.
 Thus a non-zero $A_{FB}$ means new physics, and new physics effects would be more obvious if $A_{FB}$ reached its peak value at $\xi=\frac{\pi}{2}$.

\subsection{$gg\to H\to\gamma Z \to \gamma \ell^-\ell^+ $ process}
 In this section, we firstly introduce the helicity amplitudes in spinor helicity formalism, then discuss their
 parity relations.
\subsubsection{Amplitudes in spinor helicity formalism}

\begin{figure}[htbp]
\begin{center}
\includegraphics[width=0.4\textwidth]{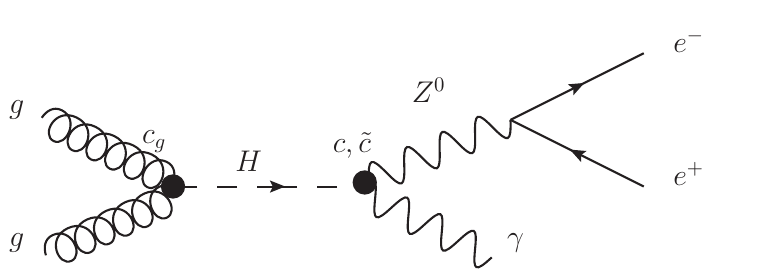}
\end{center}
 \setlength{\abovecaptionskip}{-0.5cm}
\caption{\it The Feynman diagram of the process $gg\to H \to \gamma Z \to \gamma \ell^- \ell^+ $. The $c_g$ and $c,~\tilde{c}$ factors represent the $Hgg$ and $H\gamma Z$ effective couplings respectively.  }
\label{fig:ggHgammaZ}
\end{figure}

Fig.~\ref{fig:ggHgammaZ} shows the Feynman diagram of the process $gg\to H \to \gamma Z \to \gamma \ell^- \ell^+ $ as described by effective couplings in Eq.~\eqref{eqn:HEF}. The helicity amplitude is written into three parts,
\beq
\mathcal{
A}_H(1^{h_1}_g,2^{h_2}_g,3^{h_3}_\gamma,4^{h_4}_{\ell^-},5^{h_5}_{\ell^+})
=\mathcal{A}^{gg\to H}(1^{h_1}_g,2^{h_2}_g)\times \frac{i P_H(s_{12})}{s_{12}}
\times \mathcal{A}^{H\to\gamma Z\to \gamma \ell^- \ell^+}(3^{h_3}_\gamma,4^{h_4}_{\ell^- },5^{h_5}_{\ell^+})~,
\label{eqn:gghzgamma}
\eeq
where~~$ P_X(s)=\frac{s}{s-M^2_X+iM_X\Gamma_X}$, $s_{12}=(p_1+p_2)^2$, and $h_i$ $(i=1\cdots5)$ are helicity labels of external particles.
$\mathcal{A}^{gg\to H}(1^{h_1}_g,2^{h_2}_g)$ is the helicity amplitude of gluon-gluon
fusion to Higgs process, and $h_1, h_2$ represent the helicities of outgoing gluons.
When writing the helicity amplitudes, we adopt the conventions used in~\cite{Dixon:1996wi, Campbell:2013una}:
\bea
&&\langle ij \rangle = \bar{u}_-(p_i) u_+(p_j),~~{[ ij ]} = \bar{u}_+(p_i) u_-(p_j), \nonumber\\
&&\langle ij \rangle[ ji ] = 2 p_i \cdot p_j,~~ s_{ij} = (p_i+p_j)^2, \nonumber \\
&& \epsilon^{\pm}(p_i,q)=\pm\frac{\langle q^{\mp}|\gamma^{\mu}|p_i^{\mp}\rangle}{\sqrt{2}\langle q^{\mp}|p_i^{\pm}\rangle},
\eea
where q is the reference momentum, $\epsilon^{\pm}(p_i,q)$ is for outgoing gluons. Then we have
 \bea
 \mathcal{A}^{gg\to H}(1^{+}_g,2^{+}_g)&=&\frac{2c_g}{v}[12]^2~, \nonumber\\
 \mathcal{A}^{gg\to H}(1^{-}_g,2^{-}_g)&=&\frac{2c_g}{v}\langle12\rangle^2~.
 \label{eqn:ggh}
\eea
To keep the $ggH$ coupling consistent with SM, we make
\beq
\frac{c_g}{v}=1/2\sum_f\frac{\delta^{a b}}{2}\frac{i}{16\pi^2}g^2_s4e
\frac{m_f^2}{2M_W s_W}\frac{1}{M^2_H}(2+s_{12}(1-\tau_H)C^{\gamma\gamma}_0(m_f^2))~,
\label{eqn::FggH}
\eeq
where $a, b=1,...,8$ are $SU(3)_c$ adjoint representation indices for the gluons,
$\tau_H=4m_f^2/M^2_{H}$, and the $C^{\gamma\gamma}_{0}(m^2)$ function is Passarino-Veltman three-point scalar functions \cite{Passarino:1978jh}. More details about Eq.\eqref{eqn::FggH} is given in Apendix \ref{app:c0c2}.

For all the other helicity amplitudes in this paper, we keep the convention that the momenta of external particles is outgoing. After embedding the CP violation phase $\xi$, the helicity amplitudes of $H\to\gamma Z\to \gamma \ell^- \ell^+$ are
\bea
\mathcal{A}^{H\to\gamma Z\to \gamma \ell^- \ell^+}(3^{+}_\gamma,4^{-}_{\ell^- },5^{+}_{\ell^+})
&=&2\frac{c}{v \cos\xi  }e^{-i\xi}\times \frac{P_{Z}(s_{45})}{s_{45}}\frac{\langle 45\rangle [35]^2}{\sqrt{2}} \times (el_e)
\nonumber\\
\mathcal{A}^{H\to\gamma Z\to \gamma \ell^- \ell^+}(3^{+}_\gamma,4^{+}_{\ell^- },5^{-}_{\ell^+})
&=&2\frac{c}{v\cos\xi }e^{-i\xi}\times \frac{P_{Z}(s_{45})}{s_{45}}\frac{\langle 45\rangle [34]^2}{\sqrt{2}} \times (-er_e)
\nonumber\\
\mathcal{A}^{H\to\gamma Z\to \gamma \ell^- \ell^+}(3^{-}_\gamma,4^{-}_{\ell^- },5^{+}_{\ell^+})
&=&2\frac{c}{v\cos\xi}e^{i\xi}\times \frac{P_{Z}(s_{45})}{s_{45}}\frac{[45]\langle 34\rangle^2}{\sqrt{2}} \times (el_e)
\nonumber\\
\mathcal{A}^{H\to\gamma Z\to \gamma \ell^- \ell^+}(3^{-}_\gamma,4^{+}_{\ell^- },5^{-}_{\ell^+})
&=&2\frac{c}{v\cos\xi }e^{i\xi}\times \frac{P_{Z}(s_{45})}{s_{45}}\frac{[45]\langle 35\rangle^2}{\sqrt{2}} \times (-er_e)~,
\label{eqn:ahzgamma}
\eea
where $s_{45}=(p_4+p_5)^2$, $l_e=v_f+a_f=\frac{-1+2s^2_W}{2s_Wc_W}$ and
 $r_e=v_f-a_f=\frac{2s^2_W}{2s_Wc_W}$. $l_e$ and $r_e$ are the left-hand and right-hand couplings of $Z$ boson
to leptons. We use the convention that $\epsilon_\mu(p)/\epsilon^\ast_\mu(p)$ for outgoing/incomng photons.

According to Eq.\eqref{eqn:ahzgamma}, the total cross section is proportional to $|\frac{c}{\cos\xi}|^2=c^2+\ctilde^2$, which could be fixed by the signal strength measured in future experiments. Even though, the phase of $\frac{c}{\cos\xi}$ is still unknown, which could affect the interference. We make a simple assumption 
that the phase of $\frac{c}{\cos\xi}$ is equal to that from SM at leading order. So  
\beq
|\frac{c}{\cos\xi}|^2=c^2+\ctilde^2=\mu_{SM}c^2_{SM}~~,
~~\frac{c}{\cos\xi}=\sqrt{\mu_{SM}}c_{SM},
\label{eqn:SMasum}
\eeq
where $\mu_{SM}$ is the ratio of experimental signal strength to SM expectation and we assume $\mu_{SM}=1$,  
$c_{SM}$ is the $H\gamma Z$ effective coupling in SM
from the triangle loop diagrams induced by fermions and W boson, which is given by
\beq
\frac{c_{SM}}{v}=1/2(F^{HZ\gamma}_f+F^{HZ\gamma}_W).
\eeq
According to ~\cite{Djouadi:1996yq}
\beq
F^{HZ\gamma}_f=\sum_f N_c\frac{i}{16\pi^2}v_f Q_f 8e^3\frac{m_f^2}{2M_W s_W}
(C^{\gamma Z}_0(m_f^2)+4C^{\gamma Z}_2(m^2_f)),
\label{eqn::FHZgamma_f}
\eeq
\beq
F^{HZ\gamma}_W=\frac{i}{16\pi^2}\frac{e^3}{M_Ws_W}
M^2_Z\cot{\theta_W}[\frac{2M^2_{H}}{M^2_W}(1-2c^2_W)C^{\gamma Z}_2(M^2_W)+4(1-6c^2_W)C^{\gamma Z}_2(M^2_W)
+4(1-4c^2_W)C^{\gamma Z}_0(M^2_W)]~,
\label{eqn::FHZgamma_w}
\eeq
 where $v_f=\frac{I^3_f-2Q_fs^2_W}{2s_Wc_W}$, $I^3_f=\pm\frac{1}{2}$, $s_W=\sin\theta_W$, $c_W=\cos\theta_W$ with $\theta_W$ being the Weinberg angle, and the $C^{\gamma Z}_{0,2}(m^2)$ functions are Passarino-Veltman three-point scalar functions \cite{Passarino:1978jh} as given in Appendix \ref{app:c0c2}.

\subsubsection{Parity relation}

The $2\to 3$ process could be factorized into a $2\to 2$ process times $1\to 2$ process,
\beq
\mathcal{A}_H(1^{h_1}_g,2^{h_2}_g,3^{h_3}_\gamma,4^{h_4}_{\ell^-},5^{h_5}_{\ell^+})
=\mathcal{A}_H(1^{h_1}_g,2^{h_2}_g,3^{h_3}_\gamma,45^{\kappa}_Z)\times
\frac{i P_{Z}(s_{45})}{s_{45}}\times
\mathcal{A}(45^{-\kappa}_Z,4^{h_4}_{\ell^-},5^{h_5}_{\ell^+}),
\eeq
where $45$ represents the $Z$ momentum with $p_{45}= p_4+p_5$. As an incoming leg with helicity $\kappa$ is equivalent to an outgoing leg with flipped helicity $-\kappa$, we use $\mathcal{A}(45^{-\kappa}_Z,4^{h_4}_{\ell^-},5^{h_5}_{\ell^+})$ for the $1\to2$ amplitude where the external momenta is considered outgoing.

According to Eq.s~\eqref{eqn:gghzgamma},~\eqref{eqn:ggh},~\eqref{eqn:ahzgamma} and under the assumption of Eq.\eqref{eqn:SMasum}, the $\xi$ dependent part could be extracted out as $e^{-i\kappa\xi }$, and the remaining part is the same as in the SM case. In $2\to 2$ process, we could write
\beq
\mathcal{A}_H(1^{h_1}_g,2^{h_2}_g,3^{h_3}_\gamma,45^{\kappa}_Z)
=\mathcal{A}^{SM}_H(1^{h_1}_g,2^{h_2}_g,3^{h_3}_\gamma,45^{\kappa}_Z)\times e^{-i\kappa\xi }.
\eeq
$\mathcal{A}^{SM}_H(1^{h_1}_g,2^{h_2}_g,3^{h_3}_\gamma,45^{\kappa}_Z)$ is propagated by the Higgs boson and is non-zero only when
$h_1 = h_2$ and $h_3 = \kappa$. For the non-zero amplitudes, the parity relation for $\mathcal{A}^{SM}_H(1^{h_1}_g,2^{h_2}_g,3^{h_3}_\gamma,45^{\kappa}_Z)$ is~\cite{Farina:2015dua}
 \beq
[\mathcal{A}^{SM~2\to2}_H]^{-h_1 -h_2}_{-h_3 -\kappa}=
[\mathcal{A}^{SM~2\to2}_H]^{h_1 h_2}_{h_3 \kappa}~,
\eeq
and the parity relation for $\mathcal{A}_H(1^{h_1}_g,2^{h_2}_g,3^{h_3}_\gamma,45^{\kappa}_Z)$ is
\beq
[\mathcal{A}^{2\to2}_H]^{-h_1 -h_2}_{-h_3 -\kappa}=
[\mathcal{A}^{2\to2}_H]^{h_1 h_2}_{h_3 \kappa}
\bigg|_{\xi\leftrightarrow-\xi}~.
\label{eqn:cph22}
\eeq
$\xi$ changes sign under CP transformation and thus is a CP violation phase. This is understandable since $\xi$ is connected to pseudoscalar coupling strength.

\subsection{$gg\to \gamma Z \to \gamma \ell^- \ell^+ $ process}

\subsubsection{Helicity amplitudes}

\begin{figure}[htbp]
\begin{center}
\includegraphics[width=0.4\textwidth]{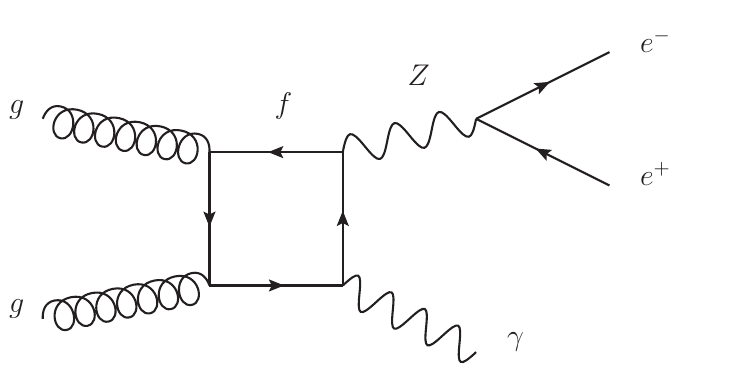}
\end{center}
 \setlength{\abovecaptionskip}{-0.5cm}
\caption{\it The Feynman diagram of the process $gg \to \gamma Z \to \gamma \ell^- \ell^+ $.}
\label{fig:gggammaZ}
\end{figure}

Fig.~\ref{fig:gggammaZ} shows the Feynman diagram of the process $gg\to \gamma Z \to \gamma \ell^- \ell^+ $.
The fermions in the loop include five light quarks. At the leading order of $\alpha_s$ expansion, only box diagrams contribute to $gg \to \gamma Z$ process \cite{Campbell:2011bn}.
The helicity amplitudes using the spinor helicity formalism are calculated in Ref.~\cite{Campbell:2011bn} and are coded in \texttt{MCFM} package.
In the following numerical analysis, we use the helicity amplitudes
in Eq.s ~(B.5)-(B.10) from Ref.~\cite{Campbell:2011bn}. We have checked the conventions carefully
to make sure the interference with $gg\to H \to \gamma Z \to \gamma \ell^- \ell^+ $ amplitudes is correct.

\subsubsection{Parity relation}

Under parity transformation the helicity amplitudes of $gg\to \gamma Z$ behave like a high-spin d-matrix function~\cite{Farina:2015dua}. The explicit expressions also support this argument~\cite{Ametller:1985di} and its parity relation is
\beq
[\mathcal{A}^{2\to2}_{box}]^{-h_1 -h_2}_{-h_3 -\kappa}=
-(-1)^\kappa[\mathcal{A}^{2\to2}_{box}]^{h_1 h_2}_{h_3 \kappa}~.
\label{eqn:cpbox22}
\eeq

\section{Kinematics and the source of $A_{FB}$ }
\subsection{The Angles }
In the helicity amplitudes, we use $p_i$ with $i=1\cdots 5$ to represent momenta of the five external legs and write the process $gg\to H\to \gamma Z \to \gamma \ell^- \ell^+ $ as
\beq
g(p_1)g(p_2) \to H(p_{12}) \to \gamma(p_3) Z(p_{45}) \to \gamma(p_3) \ell^- (p_4) \ell^+(p_5),
\eeq
where $p_{12}=p_1+p_2$, $p_{45}=p_4+p_5$.
Actually, the five momenta should satisfy energy-momentum conversation and we only
need five independent variables to character the full kinematics.
The independent variables are constructed to be the two squared invariant masses $s_{12}$ and $s_{45}$, and the three angles $\theta$, $\theta_1$ and $\phi_1$.
By contrast, in $gg\to H\to Z Z \to 4\ell$ channel, two more angles are needed
to describe another $Z$ decay plane (e.g. see Fig.~$1$ in \cite{Anderson:2013afp}).

\begin{figure}[htbp]
\begin{center}
\includegraphics[width=0.4\textwidth]{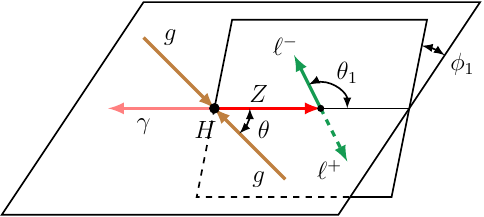}
\end{center}
\setlength{\abovecaptionskip}{-0.5cm}
\caption{\it The kinematic angles for $gg\to H\to\gamma Z \to \gamma \ell^-\ell^+ $ process.
$\theta$ is the polar angle of $Z$ boson in $H$ (or $gg$) rest frame.
$\theta_1$ is the angle of $\ell^-$ in $Z$ boson rest frame.
The z-axis of $Z$ boson rest frame is defined as the $Z$ boson production momentum direction in $H$ rest frame. $\phi_1$ is the angle between $Z$ boson production and decay planes.  }
\label{fig:hzgammakinematics}
\end{figure}

Fig.~\ref{fig:hzgammakinematics} illustrates the three angles.
$\theta\in [0,\pi]$ is the angle between $Z$ boson momentum direction and z-axis (beam direction) in $H$ rest frame. For the background process, $\theta$ is defined in the $gg$ rest frame.
 In its expression we use $-\vec{p}_3$ to represent $Z$ boson momentum direction, that is
\beq
\theta=\cos^{-1}\left(-\frac{\vec{p}_3\cdot \hat{n}_z}{|\vec{p}_3||\hat{n}_z|}\right),~\hat{n}_z=(0,0,1)~.
\eeq
$\theta_1\in [0,\pi]$ is the angle between $\ell^-$ momentum in $Z$ boson rest frame and $Z$ boson
production momentum which is obtained in $H$ rest frame. The expression for $\theta_1$ is
\beq
\theta_1=\cos^{-1}\left(-\frac{\vec{p}_3\cdot \vec{p}_4}{|\vec{p}_3||\vec{p}_4|}\right).
\eeq
 $\phi_1 \in [-\pi,\pi]$ is the angle between the $Z$ production and decay planes. We define it in the $H$ rest frame. It could also be defined in the $Z$ rest frame since any boost along the $Z$ direction won't change this angle. The expression for $\phi_1$ is
\beq
\phi_1=\frac{-\vec{ p}_{3}\cdot(\hat{n}_{prod}\times\hat{n}_{decay})}
{|\vec{p}_{3}\cdot(\hat{n}_{prod}\times\hat{n}_{decay})|}
\times\cos^{-1}(\hat{n}_{prod}\cdot\hat{n}_{decay})~,
\eeq
with $\hat{n}_{prod}$ and $\hat{n}_{decay}$ being perpendicular to the corresponding planes, which are
\bea
\hat{n}_{prod}&=&\frac{-\hat{n}_z\times \vec{p}_{3}}{|\hat{n}_z\times \vec{p}_{3}|},~\hat{n}_z=(0,0,1)~.
\nonumber \\
\hat{n}_{decay}&=&\frac{\vec{p}_4\times \vec{p}_5}{|\vec{p}_4\times \vec{p}_5|}~.
\eea

\subsection{Cross Section Factorization}

In this work we consider on-shell $Z$ boson with narrow-width approximation
for the $Z$ boson propagator, that is
$\frac{P(s_{45})}{s_{45}}\rightarrow \pi\frac{1}{M_Z\Gamma_Z}\delta(s_{45}-M^2_Z)$. The complete differential cross section is
\beq
\frac{d\hat\sigma(s_{12}, \theta; \theta_1, \phi_1) }{d(\cos\theta) d(\cos\theta_1) d\phi_1}=
\frac{(s_{12}- {M_Z}^2)}{2^{11}\pi^3s_{12}^2}\frac{\big|\mathcal{A}(s_{12},\theta;\theta_1,\phi_1,\xi)\big|^2}{M_Z\Gamma_Z}~,
\label{eqn:dcrosssection}
\eeq
where
\bea
\big|\mathcal{A}(s_{12},\theta;\theta_1,\phi_1,\xi)\big|^2
&=&\sum_{h_i}\left|\sum_{\kappa=+,0,-}
\mathcal{A}(1^{h_1}_g,2^{h_2}_g,3^{h_3}_\gamma,45^{\kappa}_Z)
\mathcal{A}(45^{-\kappa}_Z,4^{h_4}_{\ell^-},5^{h_5}_{\ell^+})\right|^2 \nonumber \\
&=&
\sum_{h_i}\left|\sum_{\kappa=+,0,-}
[\mathcal{
A}_H^{2\to2}+\mathcal{
A}_{box}^{2\to2}]^{h_1 h_2}_{h_3 \kappa}(s_{12},\theta,\xi)
[\mathcal{
A}^{1\to2}]^{-\kappa}_{h_4 h_5}(\theta_1,\phi_1)\right|^2 \nonumber \\
&=&
\sum_{\kappa,\kappa^\prime}
[\tilde{\sigma}^{2\to2}]_{\kappa \kappa^{\prime}}(s_{12},\theta,\xi)
[\tilde{\sigma}^{1\to2}]^{-\kappa -\kappa^{\prime}}(\theta_1,\phi_1)
\label{eqn:ampsquare}
\eea
with
\bea
&&[\tilde{\sigma}^{2\to2}]_{\kappa\kappa^{\prime}}(s_{12},\theta,\xi)
=\sum_{h_1,h_2,h_3}[\mathcal{
A}_H^{2\to2}+\mathcal{
A}_{box}^{2\to2}
]^{h_1h_2}_{h_3\kappa}(s_{12},\theta,\xi)[\mathcal{
A}_H^{\ast2\to2}+\mathcal{
A}_{box}^{\ast2\to2}
]^{h_1h_2}_{h_3\kappa^\prime}(s_{12},\theta,\xi)~,\nonumber\\
&&[\tilde{\sigma}^{1\to2}]^{-\kappa-\kappa^{\prime}}(\theta_1,\phi_1)
= \sum_{h_4,h_5}[\mathcal{
A}^{1\to2}]^{-\kappa}_{h_4 h_5}(\theta_1,\phi_1)
[\mathcal{
A}^{\ast1\to2}]^{-\kappa^\prime}_{h_4h_5}(\theta_1,\phi_1)~.
\label{eqn:sigma12}
\eea
The details of $\tilde{\sigma}^{1\to2}$ and $\tilde{\sigma}^{2\to2}$ are shown in the following sections and the source for $A_{FB}$ is studied afterwards. More details involving the strong phase and mass integral region will be evaluated by the end of this chapter.

\subsubsection{The $\tilde{\sigma}^{1\to2}$ contribution}

In the $Z$ rest frame, we choose
\bea
&&\epsilon^\mu(p_Z,\kappa=-)=\frac{1}{\sqrt{2}}(0,1,-i,0)\nonumber\\
&&\epsilon^\mu(p_Z,\kappa=+)=\frac{1}{\sqrt{2}}(0,-1,-i,0)\nonumber\\
&&\epsilon^\mu(p_Z,\kappa=0)=(0,0,0,1)
\eea
Then the $Z\to \ell^- \ell^+$ amplitudes are
\bea
&&[\mathcal{A}^{1\to2}]^+_{+-}(\theta_1,\phi_1)
=\frac{1}{\sqrt{2}}M_Z r_e e^{i\phi_1}(1+\cos\theta_1)\nonumber\\
&&[\mathcal{A}^{1\to2}]^+_{-+}(\theta_1,\phi_1)
=-\frac{1}{\sqrt{2}}M_Z l_e e^{i\phi_1}(1-\cos\theta_1)\nonumber\\
&&[\mathcal{A}^{1\to2}]^-_{+-}(\theta_1,\phi_1)
=\frac{1}{\sqrt{2}}M_Z r_e e^{-i\phi_1}(1-\cos\theta_1)\nonumber\\
&&[\mathcal{A}^{1\to2}]^-_{-+}(\theta_1,\phi_1)
=-\frac{1}{\sqrt{2}}M_Z l_e e^{-i\phi_1}(1+\cos\theta_1)\nonumber\\
&&[\mathcal{A}^{1\to2}]^0_{+-}(\theta_1,\phi_1)
=M_Z r_e \sin\theta_1\nonumber\\
&&[\mathcal{A}^{1\to2}]^0_{-+}(\theta_1,\phi_1)
=M_Z l_e \sin\theta_1
\eea
Thus the $[\tilde{\sigma}^{1\to2}]^{\kappa\kappa^{\prime}}$ could be written in the matrix form as
\bea
&&\left( \begin{array}{ccc}
[\tilde{\sigma}^{1\to2}]^{--} & [\tilde{\sigma}^{1\to2}]^{-0} &[\tilde{\sigma}^{1\to2}]^{-+}  \\ \nonumber
[\tilde{\sigma}^{1\to2}]^{0-} & [\tilde{\sigma}^{1\to2}]^{00} &[\tilde{\sigma}^{1\to2}]^{0+}\\ \nonumber
[\tilde{\sigma}^{1\to2}]^{+-} & [\tilde{\sigma}^{1\to2}]^{+0} &[\tilde{\sigma}^{1\to2}]^{++}\\
\end{array} \right)
\nonumber\\
&&=\frac{M^2_Z}{2}
\left( \begin{array}{ccc}
(r_e^2+l_e^2)(1+\cos^2\theta_1) & (r_e^2-l_e^2)\sqrt2\sin\theta_1e^{-i\phi_1}
& (r_e^2+l_e^2)(1-\cos^2\theta_1)e^{-i2\phi_1}\\
(r_e^2-l_e^2)\sqrt2\sin\theta_1e^{i\phi_1} & (r_e^2+l_e^2)2\sin^2\theta_1
& (r_e^2-l_e^2)\sqrt2\sin\theta_1e^{-i\phi_1}\\
(r_e^2+l_e^2)(1-\cos^2\theta_1)e^{i2\phi_1} & (r_e^2-l_e^2)\sqrt2\sin\theta_1e^{i\phi_1}
&(r_e^2+l_e^2)(1+\cos^2\theta_1)
\end{array} \right)
\nonumber
\eea
\beq
+\frac{M^2_Z}{2}\cos\theta_1
\left( \begin{array}{ccc}
-2(r_e^2-l_e^2) & -(r_e^2+l_e^2)\sqrt2\sin\theta_1e^{-i\phi_1}
& 0 \\
-(r_e^2+l_e^2)\sqrt2\sin\theta_1e^{i\phi_1} & 0
& (r_e^2+l_e^2)\sqrt2\sin\theta_1e^{-i\phi_1}\\
0 & (r_e^2+l_e^2)\sqrt2\sin\theta_1e^{i\phi_1} & 2(r_e^2-l_e^2)
\end{array} \right)
\label{eqn:matrixthetaphi}
\eeq

We split up the $[\tilde{\sigma}^{1\to2}]^{\kappa\kappa^{\prime}}$ matrix into $\cos\theta$ symmetric and asymmetric components. Notice that when $\kappa\ne\kappa^\prime$, the $[\tilde{\sigma}^{1\to2}]^{\kappa\kappa^{\prime}}$ terms depend on $\phi_1$ and
have zero contribution to the cross section after $\phi_1$ integral (from $-\pi$ to $\pi$).
That is
\beq
\begin{array}{cc}
\int^\pi_{-\pi}d{\phi_1}[\tilde{\sigma}^{1\to2}]^{\kappa\kappa^\prime}
(\theta_1,\phi_1)=0,
& \kappa\ne\kappa^\prime
\end{array}
\eeq

To study the source of $A_{FB}$, after $\phi_1$ integral, we only need to focus on the $\kappa=\kappa^\prime$ case.

\subsubsection{The $\tilde{\sigma}^{2\to2}$ contribution}

By factorizing out the $\xi$ dependence in $\tilde{\sigma}^{2\to2}$, one would have
\bea
[\tilde{\sigma}^{2\to2}]_{\kappa\kappa^{\prime}}(s_{12},\theta,\xi)
&=&[\tilde{\sigma}_{H,H}^{2\to2}]_{\kappa\kappa^{\prime}}(s_{12},\theta)
+[\tilde{\sigma}_{box,box}^{2\to2}]_{\kappa\kappa^{\prime}}(s_{12},\theta)\nonumber \\
&+&[\tilde{\sigma}_{H,box}^{2\to2}]_{\kappa\kappa^{\prime}}(s_{12},\theta)
 e^{-i\kappa\xi}
+[\tilde{\sigma}_{box,H}^{2\to2}]_{\kappa\kappa^{\prime}}(s_{12},\theta)
e^{i\kappa^\prime\xi}~,
\label{eqn:sigma22}
\eea
where $[\tilde{\sigma}_{H,H}^{2\to2}]$ represents the contribution from $gg\to H\to \gamma Z$ process,
$[\tilde{\sigma}_{box,box}^{2\to2}]$ represents the contribution from $gg\to \gamma Z$ process, and $[\tilde{\sigma}_{H,box}^{2\to2}]$ represents their interference.

According to Eqs.~\eqref{eqn:cph22}, \eqref{eqn:cpbox22} and the definition of complex conjugate, we have the following identities:
\beq
[\tilde{\sigma}_{H/box,H/box}^{2\to2}]_{+,+}
=[\tilde{\sigma}_{H/box,H/box}^{2\to2}]_{-,-}
\label{eqn:cphbox22int}
\eeq
\beq
[\tilde{\sigma}_{box,H}^{2\to2}]_{\kappa\kappa^{\prime}}=
[\tilde{\sigma}_{H,box}^{2\to2}]^\ast_{\kappa\kappa^{\prime}}.
\label{eqn:cphbox22conj}
\eeq
Applying Eqs.~\eqref{eqn:cphbox22int} and \eqref{eqn:cphbox22conj} to \eqref{eqn:sigma22}, one would have
\bea
&&[\tilde{\sigma}^{2\to2}]_{++}
-[\tilde{\sigma}^{2\to2}]_{--}
= 4\operatorname{Im}[\tilde{\sigma}_{H,box}^{2\to2}]_{++}
\sin\xi \\
&&
[\tilde{\sigma}^{2\to2}]_{++}
+[\tilde{\sigma}^{2\to2}]_{--}
= 2[\tilde{\sigma}_{H,H}^{2\to2}]_{++}+
2[\tilde{\sigma}_{box,box}^{2\to2}]_{++}+
4\operatorname{Re}[\tilde{\sigma}_{H,box}^{2\to2}]_{++}
\cos\xi
\label{eqn:sigma22p} \\
&&[\tilde{\sigma}^{2\to2}]_{00}=
[\tilde{\sigma}_{box,box}^{2\to2}]_{00}~.
\label{eqn:asymmetricfactor}
\eea

\subsection {The source of $A_{FB}$ }

Firstly, we get $A_{FB}(\hat{s})$ in gluon-gluon fusion from the above differential cross sections.
Secondly, we connect it to the $A_{FB}$ in proton-proton collision through the convolution with parton distribution function.
Finally, we show the non-resonant strong phases make $A_{FB}(\hat{s})$ change sign around the resonant peak and
propose an mass integral region asymmetric around the resonant peak to enhance $A_{FB}$.

\subsubsection {$A_{FB}(\hat{s})$ in gluon-gluon fusion}
Combining Eqs.~\eqref{eqn:dcrosssection}, ~\eqref{eqn:ampsquare}, ~\eqref{eqn:matrixthetaphi} and ~\eqref{eqn:asymmetricfactor},
we could get

\bea
\frac{d\hat\sigma(s_{12}, \theta; \theta_1, \phi_1) }{d(\cos\theta_1)}
=\frac{(s_{12}- {M_Z}^2)}{2^{11}\pi^3s_{12}^2M_Z\Gamma_Z}
& \frac{M^2_Z}{2}\bigg\{
 (r^2_e+l^2_e)\int^1_{-1}d\cos\theta\int^\pi_{-\pi}d{\phi_1}([\tilde{\sigma}^{2\to2}]_{++}
+[\tilde{\sigma}^{2\to2}]_{--})(1+\cos^2\theta_1) \nonumber \\
&+ 2(r^2_e+l^2_e)\int^1_{-1}d\cos\theta\int^\pi_{-\pi}d{\phi_1}[\tilde{\sigma}^{2\to2}]_{00}\sin^2\theta_1 \nonumber\\
&+2(l^2_e-r^2_e)\int^1_{-1}d\cos\theta\int^\pi_{-\pi}d{\phi_1}([\tilde{\sigma}^{2\to2}]_{++}
-[\tilde{\sigma}^{2\to2}]_{--})\cos\theta_1 \bigg\}~.
\eea

The forward-backward asymmetry in gluon-gluon fusion is
\bea
A_{FB}(\hat{s})
&=&\frac{N_F(\hat{s})-N_B(\hat{s})}{N_F(\hat{s})+N_B(\hat{s})}\\
&=&\frac{(\int^1_0-\int^0_{-1})d\cos{\theta_1}
\int^1_{-1}d\cos\theta\int^\pi_{-\pi}d{\phi_1}
\frac{d\hat\sigma(s_{12}, \theta; \theta_1, \phi_1) }{d(\cos\theta) d(\cos\theta_1) d\phi_1}}
{(\int^1_{-1})d\cos{\theta_1}
\int^1_{-1}d\cos\theta\int^\pi_{-\pi}d{\phi_1}
\frac{d\hat\sigma(s_{12}, \theta; \theta_1, \phi_1) }{d(\cos\theta) d(\cos\theta_1) d\phi_1}}\\
&=&
\frac{ 3(l^2_e-r^2_e)\int^1_{-1}d\cos\theta
\operatorname{Im}[\tilde{\sigma}_{H,box}^{2\to2}]_{++}
\sin\xi}
{(r^2_e+l^2_e)\int^1_{-1}d\cos\theta
(2[\tilde{\sigma}_{H,H}^{2\to2}]_{++}+
2[\tilde{\sigma}_{box,box}^{2\to2}]_{++}+
4\operatorname{Re}[\tilde{\sigma}_{H,box}^{2\to2}]_{++}
\cos\xi+[\tilde{\sigma}^{2\to2}]_{00})}~,
\label{eqn:afb}
\eea
where $\hat{s}=s_{12}$.
The denominator of $A_{FB}(\hat{s})$ includes signal and background cross sections as well as the interference part which is proportional to $\cos\xi$ .
The numerator is proportional to $\sin\xi$. In the SM case, where $\xi=0$, no $A_{FB}(\hat{s})$ could be observed.
When $\xi=\frac{\pi}{2}$, $A_{FB}(\hat{s})$ is non-zero and reaches maximum value. The detailed structure of $A_{FB}(\hat{s})$ depends on both the imaginary and real parts of $[\tilde{\sigma}_{H,box}^{2\to2}]_{++}$.

\subsubsection{$A_{FB}$ in proton-proton collision}

The proton-proton differential cross section is
\beq
\frac{d\sigma_{pp\to\gamma Z\to \gamma \ell^-\ell^+}}{d(\sqrt{\hat{s}})d(\cos\theta_1)}=2\sqrt{\hat{s}}G(\hat{s})
\frac{d\hat\sigma(\hat{s},\theta_1)}{d(\cos\theta_1)}~,
\eeq
where $\sqrt{\hat{s}}=M_{\gamma Z}$, $s$ is the total hadronic center of mass energy and
$G(\hat{s})$ is gluon-gluon luminosity function written as
\beq
G(\hat{s})=\int^1_{\hat{s}/s}\frac{dx}{sx}[g(x)g(\hat{s}/(sx)]~.
\eeq

The forward-backward asymmetry in proton-proton collision is
\bea
A_{FB}
&=&\frac{N_F-N_B}{N_F+N_B}\\
&=&\frac{(\int^1_0-\int^0_{-1})d\cos{\theta_1}\int_I d\sqrt{\hat{s}}
\frac{d\sigma_{pp\to\gamma Z\to \gamma \ell^-\ell^+}}{d(\sqrt{\hat{s}})d(\cos\theta_1)}}
{(\int^1_{-1})d\cos{\theta_1}\int_I d\sqrt{\hat{s}}
\frac{d\sigma_{pp\to\gamma Z\to \gamma \ell^-\ell^+}}{d(\sqrt{\hat{s}})d(\cos\theta_1)}}~,
\eea
where $\int_I$ represents an mass region to be integrated. The integrand in the numerator is
proportional to $\operatorname{Im}[\tilde{\sigma}_{H,box}^{2\to2}]_{++}$ and we need to further study
its dependence on $\sqrt{\hat{s}}$ to search for the suitable mass integral region.

\subsubsection{Strong phase and mass integral region }

The strong phase $\psi_1$ in $gg\to H\to\gamma Z$ process has three sources: Higgs propagator, $Hgg$ vertex and  $H\gamma Z$ vertex. With its finite width, Higgs propagator provides a strong phase that is small when far away from resonance, but increase
rapidly to $\frac{\pi}{2}$ at $M_H$. The $Hgg$ and $H\gamma Z$ vertices get small strong phases ($\sim \arctan(0.01)$ or less) from bottom loop diagrams since $M_H>2M_b$. The strong phase $\psi_2$ in $gg\to\gamma Z$
process could be introduced by light quarks (with five active flavours), which may also be suppressed by light quarks' small mass. With the assumption of zero-mass limit, the same-helicity $g^{\pm}g^{\pm}\to f\bar{f}$ process is absent.

If one extracts the strong phase $\psi^\prime_1=\tan^{-1}\frac{-M_H\Gamma_H}{s-M^2_H}$ from Higgs resonance, the other strong phases ( non-resonant strong phases ) depend more smoothly on $\sqrt{\hat{s}}$.
For this reason, we write
\bea
\operatorname{Im}[\tilde{\sigma}_{H,box}^{2\to2}]_{++}&\propto&
-\operatorname{Im}(\mathcal{A}^{gg\to H}\mathcal{A}^{H\to\gamma Z}
\mathcal{A}_{box}^{\ast gg\to \gamma Z})
\frac{\hat{s}-M^2_H}{(\hat{s}-M^2_H)^2+M_H^2\Gamma_H^2}\nonumber\\
&+&\operatorname{Re}(\mathcal{A}^{gg\to H}\mathcal{A}^{H\to\gamma Z}
\mathcal{A}_{box}^{\ast gg\to \gamma Z})
\frac{M_H\Gamma_H}{(\hat{s}-M^2_H)^2+M_H^2\Gamma_H^2}~
\label{eqn:sigma22pim}
\eea
and define a new strong phase by $\psi=\psi_1-\psi^\prime_1-\psi_2$ which is
\beq
\psi = \tan^{-1}\frac{\operatorname{Im}(\mathcal{A}^{gg\to H}\mathcal{A}^{H\to\gamma Z}
\mathcal{A}_{box}^{\ast gg\to \gamma Z})}
{\operatorname{Re}(\mathcal{A}^{gg\to H}\mathcal{A}^{H\to\gamma Z}
\mathcal{A}_{box}^{\ast gg\to \gamma Z})}~.
\eeq
From above expressions of $\psi$ and $\psi^\prime_1$, one can rewrite Eq.~\eqref{eqn:sigma22pim} and get
$\operatorname{Im}[\tilde{\sigma}_{H,box}^{2\to2}]_{++}\propto\sin(\psi_1-\psi_2)$, which is consistent with Eq.~\eqref{eqn:A}.

\begin{figure}[htbp]
\begin{center}
\includegraphics[width=0.4\textwidth]{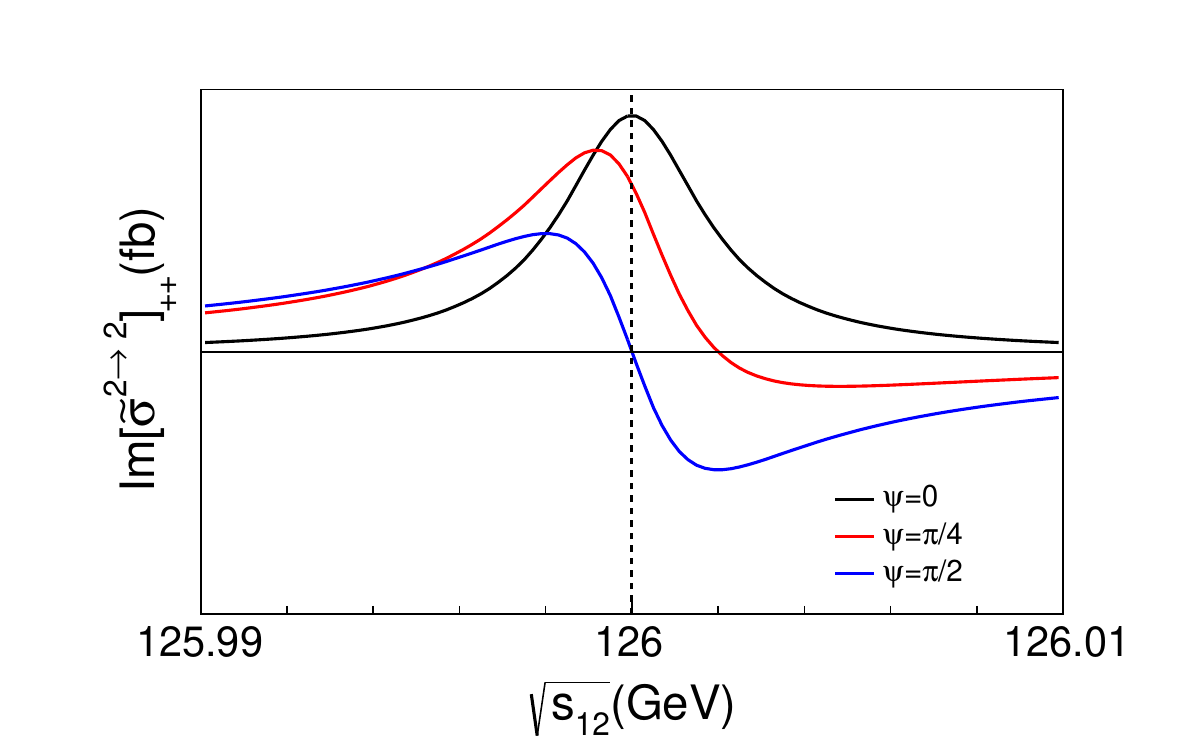}
\setlength{\abovecaptionskip}{-0.4cm}
\caption{\it $\operatorname{Im}[\tilde{\sigma}_{H,box}^{2\to2}]_{++}$ versus $\sqrt{\hat{s}}$ for different $\psi$ values.
 $M_H=126$~GeV, and $\Gamma_H=4.3$~MeV.}
\label{fig:funcIntpsi}
\end{center}
\end{figure}

Fig.~\ref{fig:funcIntpsi} shows $\operatorname{Im}[\tilde{\sigma}_{H,box}^{2\to2}]_{++}$ versus $\sqrt{\hat{s}}$ for different $\psi$ values. The values of $M_H$ and $\Gamma_H$ are set as $126$~GeV and $4.3$~MeV respectively.
 If $\psi=0$, it is the black line that is symmetric around $M_H$, which is positive through the whole resonance mass region; if $\psi$ is non-zero, $\operatorname{Im}[\tilde{\sigma}_{H,box}^{2\to2}]_{++}$ changes sign around resonant peak. For $\psi=\frac{\pi}{2}$ it changes sign at resonant peak;  for $\psi=\frac{\pi}{4}$ it changes sign when $\sqrt{\hat{s}}\approx M_H+\frac{\Gamma_H}{2}$. The asymmetric line has a long flat tail when $\sqrt{\hat{s}}$ is a few GeV far away from the resonant peak, but the symmetric line drops more rapidly. After integrated by half region below $M_H$, for example [124, 126]~GeV, the asymmetric line gets 4 times larger of the integrated value than the symmetric one. For $\psi=\frac{\pi}{4}$ , integral over [124, 126]~GeV is about 3 times of integral over [124, 128]~GeV.

  As $A_{FB}$ is proportional to the integrand of $\operatorname{Im}[\tilde{\sigma}_{H,box}^{2\to2}]_{++}$, choosing a mass integral region in which $\operatorname{Im}[\tilde{\sigma}_{H,box}^{2\to2}]_{++}$ value always has the same sign is the key factor to enhance $A_{FB}$. When resonant width is very small, a mass region one-side below or above resonant peak could fulfill this criterion and supply a relatively large $A_{FB}$. In the following simulation we make a comparison for $A_{FB}$ values between one-side and symmetric mass integral regions.

\section{Simulation}

The simulations to quantify interference effects and the value of $A_{FB}$ is preformed using \texttt{MCFM} package. We adopt the amplitudes for $gg\to \gamma Z\to\gamma \ell^- \ell^+$ process from \texttt{MCFM} and add amplitudes for $gg\to H\to \gamma Z\to\gamma \ell^- \ell^+$ as described in previous sections. The simulations are generated for a proton-proton collider with $\sqrt s=14$~TeV. The final state photon is required to have $p^{\gamma}_T>20$~GeV and $|\eta^{\gamma}|<2.5$. The $\ell^-\ell^+$
invariant mass is set to be in the $Z$ boson mass region which is $M_{\ell^-\ell^+}\in$ [66, 116]~GeV.

\begin{figure}[htbp]
\begin{center}
\begin{minipage}[t]{0.4\textwidth}
\includegraphics[width=1.0\textwidth]{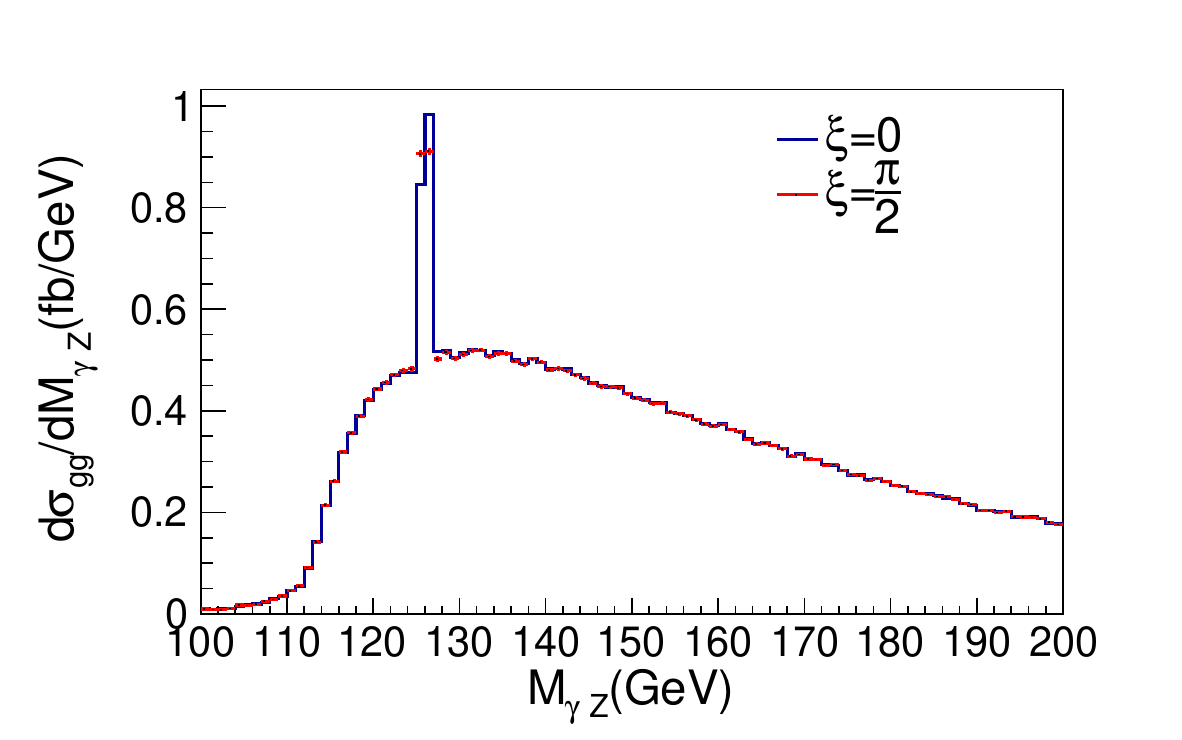}
\end{minipage}
\begin{minipage}[t]{0.4\textwidth}
\includegraphics[width=1.0\textwidth]{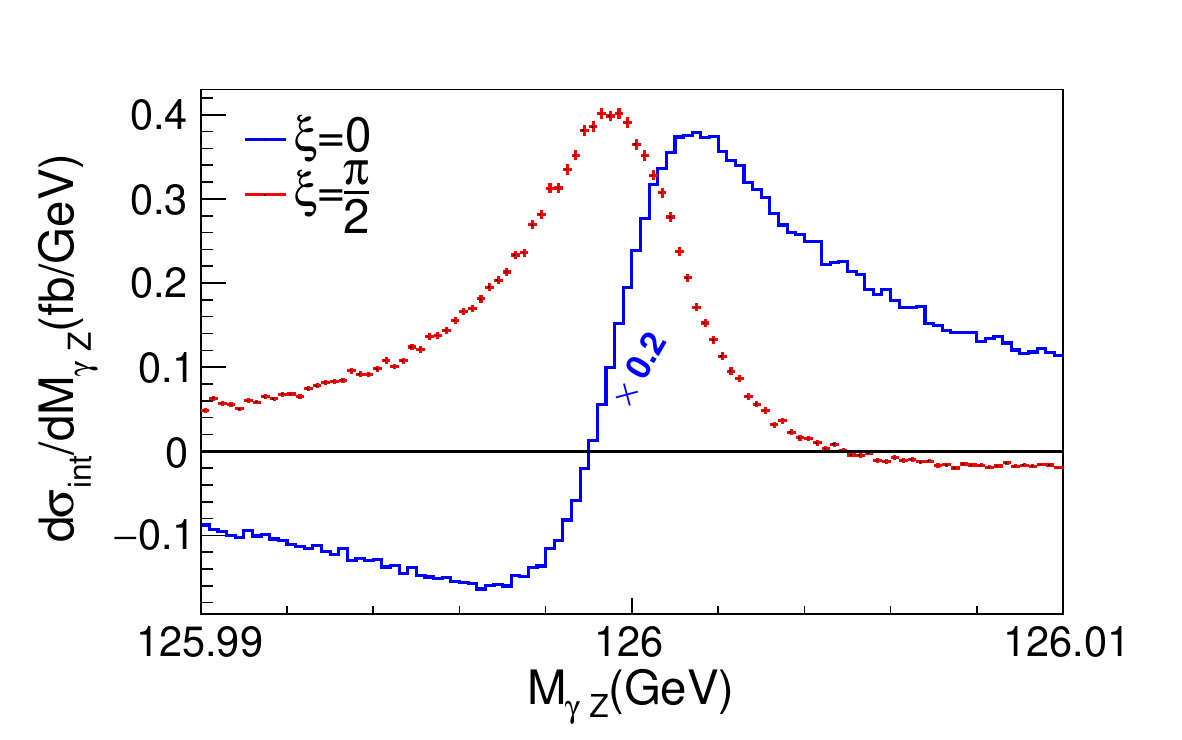}
\end{minipage}
\setlength{\abovecaptionskip}{-0.45cm}
\caption{\it Left panel: The $M_{\gamma Z}$ differential cross section for $\xi=0$ (blue histogram) and $\xi=\frac{\pi}{2}$
(red histogram with error bars) cases. Right panel: The $M_{\gamma Z}$ differential cross section of only the interference part ($\propto|\mathcal{A}_H+\mathcal{A}_{box}|^2-|\mathcal{A}_H|^2-|\mathcal{A}_{box}|^2$) integrate over $\cos\theta_1$ from 0 to 1. The blue histogram is scaled by $\times 0.2$. }
\label{fig:dsigmagammaz}
\end{center}
\end{figure}

\begin{figure}[htbp]
\begin{center}
\includegraphics[width=0.4\textwidth]{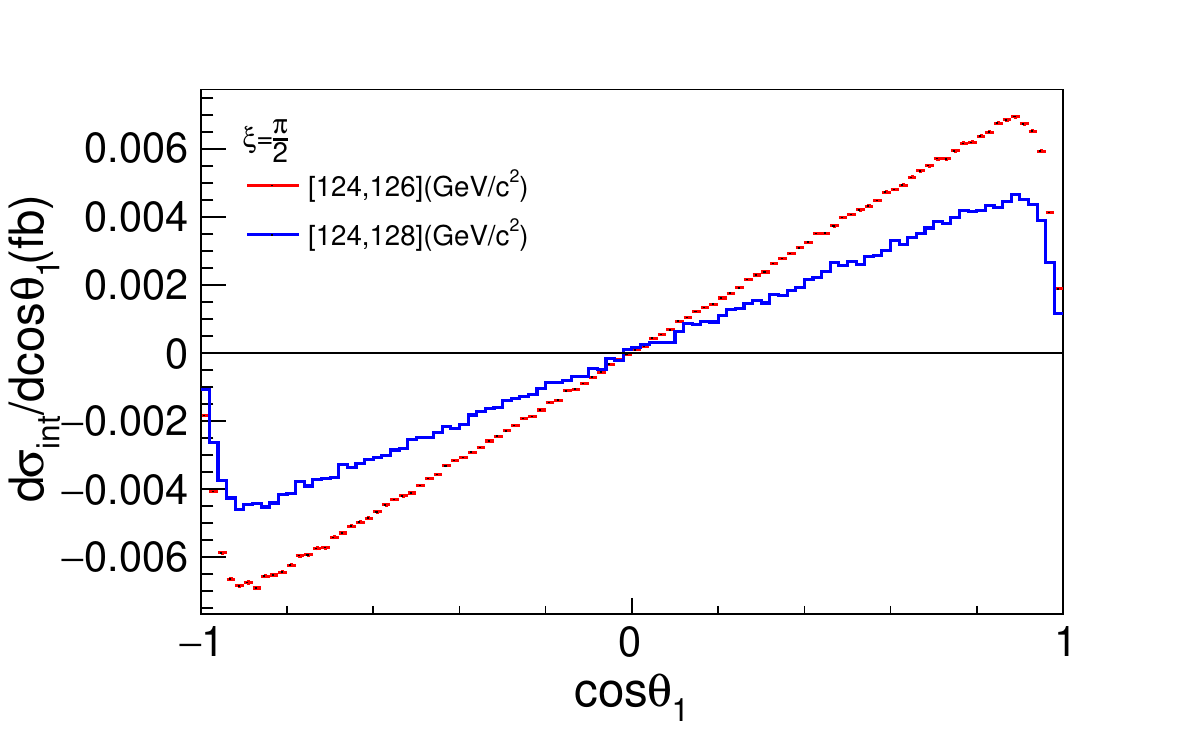}
\setlength{\abovecaptionskip}{-0.45cm}
\caption{\it $d\sigma_{int}/d\cos{\theta_1}$ versus $\cos{\theta_1}$ when $\xi=\frac{\pi}{2}$ for $[124,126]$~GeV integral region (red histogram with error bars) and $[124,128]$~GeV integral region (blue histogram).}
\label{fig:asymmetricfactor}
\end{center}
\end{figure}

Fig.~\ref{fig:dsigmagammaz} left panel shows the fiducial differential cross section for $gg\to \gamma Z\to\gamma \ell^- \ell^+$ process including the $gg\to H \to\gamma Z\to\gamma \ell^- \ell^+$ process and their interference part.
 The peak at $M_H=126$~GeV is the Higgs resonance and the width of the peak is about $4.3$~MeV. In the high mass region, the cross section decreases slowly as $M_{\gamma Z}$ increases. The blue and red histograms represent $\xi=0$ and $\xi=\frac{\pi}{2}$ cases respectively. They have small difference at the resonant region caused by interference effects. Fig.~\ref{fig:dsigmagammaz} right panel shows only the interference contribution, which is calculated by $|\mathcal{A}_H+\mathcal{A}_{box}|^2-|\mathcal{A}_H|^2-|\mathcal{A}_{box}|^2$. For both cases when $\xi=0$ or $\frac{\pi}{2}$, we integrate only half region of $\cos\theta_1$ (from 0 to 1) to keep the contribution from $\cos\theta_1$-odd terms. This treatment is based on the fact that when $\xi=\frac{\pi}{2}$ the $\cos\theta_1$ distribution of the interference part is asymmetric (see Fig. \ref{fig:asymmetricfactor}). When $\xi=0$, the interference contribution is proportional to $\operatorname{Re}[\tilde{\sigma}_{H,box}^{2\to2}]_{++}$ (the blue histogram), which also consists of large asymmetric contribution. When $\xi=\frac{\pi}{2}$, the interference contribution is proportional to $\operatorname{Im}[\tilde{\sigma}_{H,box}^{2\to2}]_{++}$ (the red histogram with error bars). From the shape of the red histogram, the peak position is shifted to left by about $1$~MeV, and the cross section reaches to zero at about $M_{\gamma Z}= 126.004~GeV$. It corresponds to $\tan\psi\sim0.5$. The integral over only half region below $M_H$ could increase the numerator of $A_{FB}$ while decrease the denominator by half. In the following analysis, we preform the integral over the lower half region of $M_{\gamma Z}$ to study the enhanced $A_{FB}$ effect.

Fig.~\ref{fig:asymmetricfactor} shows the differential cross section of $\cos\theta_1$ (with $\xi=\frac{\pi}{2}$). The slope represents the numerator value of $A_{FB}$. When integrating over the asymmetric region ($[124, 126]$~GeV) around the $M_H$, the value of the slope is about 0.008. By contrast, when integrating over the symmetric region ($[124, 128]$~GeV), the value of the slope is about 0.005. Together with the total cross section, $A_{FB}$ values under different integral regions are summarized in Table~\ref{table:afb}.

\begin{table}[htbp]
\begin{center}
\caption{\it $A_{FB}$ values under different integral mass regions.}
\begin{tabular}{c|c|c|c}
\hline
Integral mass region(GeV) & $A_{FB}$ numerator (fb) & $A_{FB}$ denomenator (fb)  & $A_{FB}$ \tabularnewline
 \hline
 [124, 126] &0.008  &1.4  &  $\sim0.57\%$   \tabularnewline
 \hline
[124, 128] &0.005  &2.8  &  $\sim0.18\%$    \tabularnewline
\hline
\end{tabular}
\label{table:afb}
\end{center}
\end{table}

The quadrant-type asymmetry $\Sigma_{\phi_1}$ defined on $\phi_1$ oscillation is another CP observable in $H\to\gamma Z$ process~\cite{Farina:2015dua}. Ref.~\cite{Farina:2015dua} shows that it is about $\frac{2}{\pi}\times(-0.84)\times 10^{-3}\sim -0.05\%$ if integrate over [124,128]~GeV region, while $A_{FB}$ is about $0.18\%$ in [124,128]~GeV region. The $\Sigma_{\phi_1}$ could also be enhanced when integrating over the half resonance region.

Theoretically when the non-resonant strong phase is neither zero nor $\pi/2$ the interference could be considered to have two parts: one asymmetric part and one symmetric part, just corresponding to the first term and second term in Eq.~\ref{eqn:sigma22pim}, or corresponding to the blue line and black line in Fig.~\ref{fig:funcIntpsi}.
When integrating over the whole resonance region, only the symmetric part of interference contributes to $A_{FB}$; when integrating over the half resonance region, 
both symmetric part and asymmetric part contribute. That is why the $A_{FB}$ is enhanced when choosing a half resonance region. 
 From this viewpoint, according to the values of $A_{FB}$ numerator in Table.~\ref{table:afb},
we could estimate that the contribution from asymmetric part of interference  
is about 2 times of the contribution from symmetric part when integrating over 
[124, 126]~GeV.  

In experiment, limited mass resolution will smear a theoretical sharp resonance peak to a wide bump. The symmetric part of the $M_{\gamma Z}$ differential cross section from interference contribution in Fig.~\ref{fig:dsigmagammaz} will be smeared to a bump while the asymmetric part will be smeared to two opposite-sign bumps. 
The resonance peak is expanded and the region of the bump would be related to the value of mass resolution.
When integrating over a half resonance region, the 
integral of the symmetric part are nearly the same 
before or after considering limited mass resolution, while the integral of the asymmetric part 
will reveal less of the asymmetric effect with experiment data.
 This is because the two opposite-sign bumps will have some overlap near the resonance peak and partially cancel each other.             
The $A_{FB}$ from a half resonance region will be weakened by mass resolution. 
Another issue about integrating over half the resonance region is the mass uncertainty. 
The fitted mass of resonance could be used as a reference point to choose the half integral region. 
If the fitted mass had a large uncertainty, the central value could be far from the theoretical peak and the $A_{FB}$ from a half resonance region may have a large deviation from our prediction. 
In practice, 
as the integral region is already expanded by mass resolution, one needs to consider the relative size between mass uncertainty and mass resolution. 
For example, at LHC in the Higgs to diphoton decay channel ~\cite{CMS:2017rli} the recent experiment shows the mass resonance region is around [121, 131]~GeV with mass resolution of $\sim1$~GeV. On the other hand, the mass uncertainty is about 0.1~GeV which is one order of magnitude smaller than the resolution and two order of magnitude smaller than the resonance region. In this situation 
the uncertainty of $A_{FB}$ caused by mass uncertainty could be ignored when integrating over half the resonance region. 

In conclusion, it is still better to consider the integral over one side of the resonance peak. The $A_{FB}$ value would still be larger than if integrated over the whole resonance region. The simulation including the mass resolution and the resonance mass uncertainty is beyond the scope of this paper. We will use $0.57\%$ from Table~\ref{table:afb} to estimate the significance in the following analysis.

The significance is estimated as the following. After the fiducial cuts of $M_{\ell^-\ell^+}\in [66, 116]$~GeV, $M_{\gamma Z}\in [124, 126]$~GeV, $p^{\gamma}_T>20$~GeV and $|\eta^\gamma|<2.5$, for $\xi=\frac{\pi}{2}$, the total cross section with interference effect of $gg\to \gamma Z\to\gamma \ell^- \ell^+$ and $gg\to H\to \gamma Z\to\gamma \ell^- \ell^+$ is $\sigma_{gg}=1.4fb$ while the cross section for background $q\bar{q}\to\gamma Z$ process is $\sigma_{q\bar{q}}=40.8fb$. According to the definition of the significance
\beq
\frac{S}{\sqrt{B}}=\frac{A_{FB}\sigma_{gg}L}{\sqrt{\sigma_{q\bar{q}}L}}
\sim\frac{A_{FB}}{0.08}\sqrt\frac{L}{3000fb^{-1}}~,
\eeq
after the high-luminosity phase of LHC (HL-LHC) reaching $3000fb^{-1}$ luminosity, the $A_{FB}$ effect from the interference contribution should be about 0.08 to reach a significance $\sim 1$. From our current model, the $A_{FB}$ effect of $0.57\%$ would still be difficult to distinguish. However, it leaves possibility at HL-LHC for new physics which could introduce both large CP-violation phases and interference effect.

\section{conclusion and discussion}

In this work we construct a model with general CP violation phase $\xi$ from $H\gamma Z$ coupling. By calculating the interference effect between $gg\to H\to\gamma Z\to\gamma \ell^-\ell^+$ and $gg\to\gamma Z\to\gamma \ell^-\ell^+$ processes, we confirm that the forward-backward asymmetry $A_{FB}$ of charged leptons in the $Z$ rest frame is a CP-violation observable, and is proportional to $\sin\xi$. We analyze the impact of several non-zero strong phases which is also a key factor to determine the value of $A_{FB}$. By studying the shape of the integrand, we propose to do integral of $M_{\gamma Z}$ over half of the resonant mass region to enhance $A_{FB}$. After detailed simulations using modified \texttt{MCFM}, we estimate the $A_{FB}$ could reach about $0.6\%$. After considering the huge amount of background process, the significance is relatively small and hard to be observed at the HL-LHC. More detailed studies involving non-zero strong phases and mass regions of $M_{\gamma Z}$ could be preformed under similar frameworks. The analysis also reveals that new physics with large CP-violation phases may not be easily ruled out when searching for forward-backward asymmetry at the LHC.

\begin{acknowledgments}
X.W. thanks Youkai Wang for valuable discussions about the cross section factorization and $A_{FB}$ definition.
We thank Yandong Liu and Chih-Hao Fu for helpful discussions.  X.W. is supported by the National Science Foundation of China under Grant No. 11405102. The work was also supported in part by the National Science Foundation of China under Grant No. 11635001 and No. 11375014.

\end{acknowledgments}

\appendix
\section{Definition of Passarino-Veltman three-point scalar functions}
\label{app:c0c2}
 For the Higgs production and decay processes, the Passarino-Veltman three-point scalar functions $C^{\gamma \gamma}_{0}(m^2)$, $C^{\gamma Z}_{0}(m^2)$ and $C^{\gamma Z}_{2}(m^2)$ have simple forms in terms of $\tau_Z=4m^2/M^2_Z$ and $\tau_H=4m^2/M^2_{H}$~:
\bea
4 m^2 C^{\gamma \gamma}_0(m^2) &=&  2\tau_H f(\tau_H)~,
\label{eqn:c0gammagamma}
\\
4 m^2 C^{\gamma Z}_0(m^2) &=& - \frac{2\tau_Z \tau_H}{\tau_Z-\tau_H}
\left[f(\tau_Z)-f(\tau_H)\right]~,
\label{eqn:c0}
\\
4 m^2 C^{\gamma Z}_2(m^2) &=& \frac{\tau_Z \tau_H}{2(\tau_Z-\tau_H)}
+\frac{\tau_Z \tau_H^2}{2(\tau_Z-\tau_H)^2}
\Big( \tau_Z \left[f(\tau_Z)-f(\tau_H)\right]
  + 2 \left[g(\tau_Z)-g(\tau_H)\right] \Big)~
\eea

with the functions $f$ and $g$ are defined by 
\begin{equation}
f(\tau) = \left\{ \begin{array}{ll}
{\rm arcsin}^2 \sqrt{1/\tau} & \tau \geq 1 \\
-\frac{1}{4} \left[ \log \frac{1 + \sqrt{1-\tau } }
{1 - \sqrt{1-\tau} } - i \pi \right]^2 \ \ \ & \tau <1~
\end{array} \right.
\end{equation}
\begin{equation}
g(\tau) = \left\{ \begin{array}{ll}
\sqrt{\tau-1} \ {\rm arcsin} \sqrt{1/\tau} & \tau \geq 1 \\
\frac{1}{2} \sqrt{1-\tau} \left[ \log \frac{1 + \sqrt{1-\tau } }
{1 - \sqrt{1-\tau} } - i \pi \right] \ \ \ & \tau <1~
\end{array} \right.
\end{equation}

\bibliographystyle{apsrev}

\begin{thebibliography}{10}

\bibitem{Gavela:1993ts}
M.~B. Gavela, P.~Hernandez, J.~Orloff, and O.~Pene,
\newblock Mod. Phys. Lett. {\bf A9}, 795 (1994), arXiv:hep-ph/9312215.

\bibitem{Sakharov:1967dj}
A.~D. Sakharov,
\newblock Pisma Zh. Eksp. Teor. Fiz. {\bf 5}, 32 (1967),
\newblock [Usp. Fiz. Nauk161,61(1991)].

\bibitem{Dekens:2013zca}
W.~Dekens and J.~de~Vries,
\newblock JHEP {\bf 05}, 149 (2013), arXiv:1303.3156.

\bibitem{Inoue:2014nva}
S.~Inoue, M.~J. Ramsey-Musolf, and Y.~Zhang,
\newblock Phys. Rev. {\bf D89}, 115023 (2014), arXiv:1403.4257.


\bibitem{Chen:2014gka}
Y.~Chen, R.~Harnik, and R.~Vega-Morales,
\newblock Phys. Rev. Lett. {\bf 113}, 191801 (2014), arXiv:1404.1336.


\bibitem{Khachatryan:2014kca}
CMS, V.~Khachatryan {\em et~al.},
\newblock Phys. Rev. {\bf D92}, 012004 (2015), arXiv:1411.3441.

\bibitem{Chen:2014ona}
Y.~Chen, A.~Falkowski, I.~Low, and R.~Vega-Morales,
\newblock Phys. Rev. {\bf D90}, 113006 (2014), arXiv:1405.6723.

\bibitem{Korchin:2014kha}
A.~{\relax Yu}. Korchin and V.~A. Kovalchuk,
\newblock Eur. Phys. J. {\bf C74}, 3141 (2014), arXiv:1408.0342.

\bibitem{Farina:2015dua}
M.~Farina, Y.~Grossman, and D.~J. Robinson,
\newblock Phys. Rev. {\bf D92}, 073007 (2015), arXiv:1503.06470.

\bibitem{Li:2015kxc}
G.~Li, H.-R. Wang, and S.-h. Zhu,
\newblock Phys. Rev. {\bf D93}, 055038 (2016), arXiv:1506.06453.

\bibitem{Dixon:1996wi}
L.~J. Dixon,
\newblock {Calculating scattering amplitudes efficiently},
\newblock in {\em {QCD and beyond. Proceedings, Theoretical Advanced Study
  Institute in Elementary Particle Physics, TASI-95, Boulder, USA, June 4-30,
  1995}}, pp. 539--584, 1996, arXiv:hep-ph/9601359.

\bibitem{Campbell:2013una}
J.~M. Campbell, R.~K. Ellis, and C.~Williams,
\newblock JHEP {\bf 04}, 060 (2014), arXiv:1311.3589.

\bibitem{Passarino:1978jh}
G.~Passarino and M.~J.~G. Veltman,
\newblock Nucl. Phys. {\bf B160}, 151 (1979).

\bibitem{Djouadi:1996yq}
A.~Djouadi, V.~Driesen, W.~Hollik, and A.~Kraft,
\newblock Eur. Phys. J. {\bf C1}, 163 (1998), arXiv:hep-ph/9701342.

\bibitem{Campbell:2011bn}
J.~M. Campbell, R.~K. Ellis, and C.~Williams,
\newblock JHEP {\bf 07}, 018 (2011), arXiv:1105.0020.

\bibitem{Ametller:1985di}
L.~Ametller, E.~Gava, N.~Paver, and D.~Treleani,
\newblock Phys. Rev. {\bf D32}, 1699 (1985).

\bibitem{Anderson:2013afp}
I.~Anderson {\em et al.}, ``{Constraining anomalous HVV interactions at proton
  and lepton colliders},''
  \href{http://dx.doi.org/10.1103/PhysRevD.89.035007}{{\em Phys. Rev.} {\bf
  D89} (2014) no.~3, 035007},
\href{http://arxiv.org/abs/1309.4819}{{\tt arXiv:1309.4819 [hep-ph]}}.


\bibitem{CMS:2017rli} 
  CMS Collaboration [CMS Collaboration],
  CMS-PAS-HIG-16-040.


\end{thebibliography}

\end{document}